\documentstyle[12pt]{article}
\begin{document}
\hsize = 6.0 in
\vsize =11.7 in
\hoffset=0.1 in
\voffset=-0.5 in
\baselineskip=20pt
\newcommand{\ghat}{{\hat{g}}}
\newcommand{\Rhat}{{\hat{R}}}
\newcommand{\ih}{{i\over\hbar}}
\newcommand{\Scal}{{\cal S}}
\newcommand{\fudge}{{1\over16\pi G}}
\newcommand{\tn}{\mbox{${\tilde n}$}}
\newcommand{\mg}{\mbox{${m_g}^2$}}
\newcommand{\mf}{\mbox{${m_f}^2$}}
\newcommand{\hk}{\mbox{${\hat K}$}}
\newcommand{\vk}{\mbox{${\vec k}^2$}}
\newcommand{\half}{{1\over2}}
\newcommand{\eqletter}{ \hfill (\theequation\alph{letter})}
\newcommand{\gm}{{(\Box+e^2\rho^2)}}
\newcommand{\eql}{\nonumber &\eqletter \cr
                  \addtocounter{letter}{1}}
\newcommand{\be}{\begin{equation}}
\newcommand{\ee}{\end{equation}}
\newcommand{\bea}{\begin{eqnarray}}
\newcommand{\eea}{\end{eqnarray}}
\newcommand{\beal}{\setcounter{letter}{1} \begin{eqnarray}}
\newcommand{\eeal}{\addtocounter{equation}{1} \end{eqnarray}}
\newcommand{\none}{\nonumber \\}
\newcommand{\req}[1]{Eq.(\ref{#1})}
\newcommand{\reqs}[1]{Eqs.(\ref{#1})}
\newcommand{\larrow}{\,\,\,\,\hbox to 30pt{\rightarrowfill}
\,\,\,\,}
\newcommand{\slarrow}{\,\,\,\hbox to 20pt{\rightarrowfill}
\,\,\,}
\newcommand{\bfx}{{\vec{x}}}
\newcommand{\bfy}{{\vec{y}}}
\newcommand{\zfp}{Z_{{FP}}}
\newcommand{\zf}{Z_{{F}}}
\newcommand{\zr}{Z_{{R}}}
\newcommand{\zop}{Z_{{OP}}}
\newcommand{\zekt}{Z_{EKT}}
\newcommand{\phstar}{{\varphi^\dagger}}
\newcommand{\calM}{{\cal M}}
\begin{center}
{\bf Quantum Bubble Dynamics in 2+1 Dimensional Gravity}\\
{I. Geometrodynamic Approach}\\
\vspace{15 pt}
{\it by}\\
\vspace{13 pt}
H. Zaidi${}^1$ {\it and} J. Gegenberg${}^2$\\[5pt]
  ${}^1${\it  Department of Physics}\\
{\it University of New Brunswick}\\
{\it Fredericton, New Brunswick}\\
{\it CANADA, E3B 5A3}\\
${}^2${\it Department of Mathematics and Statistics}\\
   {\it University of New Brunswick}\\
   {\it Fredericton, New Brunswick}\\
   {\it CANADA E3B 5A3}\\[5pt]
\end{center}
\vspace{40pt}
{\narrower\smallskip\noindent
{\bf Abstract} :
The Dirac quantization of a 2+1 dimensional bubble is performed.  The
bubble consists of a string forming a boundary between two regions of
space-time with distinct geometries.  The ADM constraints are solved
and the coupling to the string is introduced through the boundary
conditions.  The wave functional is obtained and the quantum
uncertainty in the radius of the ring is calculated; this uncertainty
becomes large at the Planck scale.
\vspace{40 pt}
\begin{center}
{\it May 1993}
\end{center}
\par
\vspace*{20 pt}
\noindent
UNB Technical Report 93-05

\noindent
gr-qc/9311016
\clearpage
\section{Introduction}
In recent years there has been considerable interest in 2+1
dimensional gravity, in part due to the fact that pure gravity can be
exactly quantized in 2+1 dimensions and hence provides some insight
into the more difficult problem of quantizing gravity in the
physically realistic 3+1 dimensional case.  The quantization can be
performed in the first order Witten-Ashtekar formalism
\cite{review,witten1,ash1}
, as well as in the ADM or geometrodynamic
formalism \cite{ADM}; but coupling to matter complicates the problem
\cite{gegkun1}.  The special case of point particles has been
considered in some detail \cite{carlip}.  Here we consider another
solvable model, namely a bubble consisting of a string forming the
boundary between between an outer flat spacetime and an inner region
with non-vanishing cosmological constant.  It is possible to quantize
this model non-perturbatively in both the ADM and first-order
formalisms.  In this paper we approach the quantization in a manner
which parallels the ADM treatment of the "bubble nucleation problem" in
3+1 dimensions \cite{polch}.  The restriction to spherical symmetry in the
latter makes this problem tractable.  The analogue in 2+1 dimensions
is to impose circular symmetry so that the string boundary between
the inner and outer geometries is circular.  A somewhat different but
related problem, with no distortion of the spacetime metric due to
the presence of the string, has also been treated classically; the
motion of a circular string in fixed backgrounds of various types has
been analyzed using the Einstein equations \cite{soda}.  However, in
our model, the string actively distorts the spacetime geometry by
modifying the metric in the two regions, but with the components of
the metric tensor continuous across the boundary.  These distortions
are determined by gauge fixing using the boundary conditions; thus
the coupling with the string is introduced.
\par
In the next section, a canonical form of the action is obtained and
the ADM constraints and boundary conditions are derived.  These
constraints are solved for the canonical momenta in terms of the
configuration space variables and we then proceed, in section 3, to Dirac
quantization, ignoring factor-ordering problems.
The general wave functional involves the gauge as well as the dynamical
modes.  After gauge fixing, the wave functional reduces to a wave {\it
function} of the string radius, the only dynamical mode remaining.  The
behaviour of the quantized system is analyzed in section 4.  Barrier
penetration is observed in some cases and quantum uncertainty in the
radius is calculated.  As expected, these quantum effects are large for
a string of radius of order of the Planck length.  The results and
possible extensions of the theory are discussed in section 5.  In a following
paper, the classical and quantum analysis of the first-order form of the
model is developed, including the extension to the case where the string
has non-radial modes
{}.
\section{Solution of the Constraints}
We consider a string in a 2+1 dimensional Lorentzian spacetime $M^3$
with metric $g_{ab}$.  We assume that the string is circular with
variable radius $\hat r$ and string tension $\mu$.  The string divides
$M^3$ into an inner region $r<\hat r$ with cosmological constant $\lambda\neq
0$ and an outer region $r>\hat r$ with zero cosmological constant.  The action
functional for this system is:

\be
S[g_{ab}, \hat r]=\fudge\int_{M^3}d^3x\sqrt{-g}\left[R-2\lambda\theta(\hat
r-r)\right]
-{\mu\over4\pi}\int_Bd^2y\sqrt{-h}, \label{eq:1.1}
\ee
where $G$ is the gravitational coupling constant, $g=\det{g_{ab}}$,
$R$ is the Ricci scalar, $\theta(x)$ is the usual unit step function, $B$ is
the trajectory of the string-and hence here is topologically a cylinder- and
finally $h_{ij}$, $i,j,...=0,1$ is the metric induced on $B$ by $g_{ab}$.
The $x^a$ are coordinates on $M^3$ and the $y^i$ are coordinates on $B$.
It is convenient to choose polar coordinates $x^0=t, x^1=r,
x^2=\phi$, so that $B$ is given by $r=\hat r$ and $y^0=t, y^1=\phi$.  In this
case the ADM line-element can be written:

\be
ds^2=-N^2dt^2+L^2\left(dr+Mdt\right)^2+R^2d\phi^2,\label{eq: adm}
\ee
where $N(t,r)$ is the lapse function, $M(t,r)$ is the radial component of
the shift vector and the spatial metric is of the form $diagonal(L^2(t,r),
R^2(t,r))$.  Using the standard technique, the action can be written in
(2+1) form as:
\bea
S=&\int dt \int dr {\cal L}(L,R,\hat r, \dot L, \dot R ,\dot{\hat r}, L',
R'), \\ \nonumber
{\cal L}:=&-N\left[{1\over4G}\left({R'\over L}\right)'-4G\Pi_R \Pi_L+
{\lambda\over4G}\theta(\hat r-r)LR+\delta(\hat r-r){\hat E\over2L}\right]\\
\nonumber
&-M\left[-L\Pi_L'+R'\Pi_R'+\delta(\hat r-r)\hat p\right] +\Pi_L\dot L
+\Pi_R\dot R+\hat p \dot{\hat r}\delta(\hat r-r),\label{eq:1.4}
\eea
where $L'=\partial L/\partial r$, $\dot L=\partial L/\partial t, \Pi_L=\partial
{\cal L}/\partial \dot L$, etc.; $\hat p=\partial {\cal L}/\partial\dot{\hat
r},
\hat E=(4\hat p^2+\mu^2\hat R^2 \hat L^2)^{\half}$; $\hat L=L(\hat r),$ etc.
In view of the circular symmetry, an integration over $\phi$ has been performed
above.

It is easy to to read off the fundamental Poisson brackets from the
canonical form of the action.  As indicated from the notation, $(L,\Pi_L),
(R,\Pi_R)$ and $(\hat r, \hat p)$ are canonically conjugate pairs.  The
lapse and shift, $N$ and $M$, play the role of Lagrange multipliers since
$\Pi_N=\Pi_M=0$, and enforce the constraints:

\be
{\cal H}_t:={1\over 4G}\left({R'\over L}\right)'-4G\Pi_R\Pi_L+\theta(\hat r
-r){\lambda\over 4G}LR+\delta(\hat r-r){\hat E\over 2\hat L}\approx 0,
\label{eq:1.5}
\ee

\be
{\cal H}_r:=R'\Pi_R-L\Pi_L'+\delta(\hat r- r)\hat p\approx 0. \label{eq:1.6}
\ee

The constraints have the same form as in the 3+1 dimensional case, and they can
be solved
in the same manner \cite{polch}.  Eliminating $\Pi_R$ we obtain from
\req{eq:1.5} and \ref{eq:1.6}:
\be
\calM'=\hat\rho\delta(\hat r-r),\label{eq:1.7}
\ee
where
\bea
\calM:=\left\{\begin{array}{ll}-(8G)^{-1}(R'/L)^2+2G\Pi^2_L,r>\hat r \\
-(8G)^{-1}[(R'/L)^2+\lambda R^2]+2G\Pi^2_L, r<\hat r \end{array}
\right.;\label{eq:1.8}
\eea
and
\be
\hat\rho:={\hat E\over 2\hat L^2}-4G{\hat p\hat\Pi_L\over \hat L}.
\ee
\req{eq:1.7} has the form of Gauss' law with the source density $\hat
\rho$ concentrated at $r=\hat r$.  Therefore, the parameter $\calM$ can be
interpreted as the mass.  Since $\calM'=0$ for $r\neq\hat r$, $\calM$ is
independent of $r$ in both regions.  Moreover, $\calM(0)=0$ since $R(r)$
is the radial coordinate with $R(0)=0,R'(0)=0$ and $\Pi_L=0$.
Therefore, $M=0,r < \hat{r}$, and $M(\infty) = M = \mbox{constant}$.
Hence \req{eq:1.8} can be solved for $\Pi_L$ in the following form
\bea
\Pi_L = \left\{ \begin{array}{ll}
4(G)^{-1}[(R'/L)^2 + \lambda R^2]^{1/2}, & r < \hat{r}\\
4(G)^{-1}[(R'/L)^2 + 8GM], & r > \hat{r}
\end{array} \right. \label{eq:1.10}
\eea
Moreover, \req{eq:1.6} gives
\be
\Pi_R = (L/R')\Pi'_L,\;\;\;r \neq \hat{r}. \label{eq:1.11}
\ee
Eqs. \req{eq:1.10} and \req{eq:1.11} represent the complete solution of the
constraints, in both regions.  The coupling to the string is specified
by the boundary conditions.  As in ref. 7, we require that $R(r)$ and
$L(r)$ be continuous, but the derivatives are discontinuous at $r =
\hat{r}$.  Moreover, $\Pi_L$ and $\Pi_R$ are discontinuous since the
string momentum, $\hat{p}$, also contributes to the momentum balance.
These considerations applied to \req{eq:1.5} and \req{eq:1.6}
lead to the following
boundary conditions
\be
\bigtriangleup \hat{\Pi}_L  =  - \hat{p}/ \hat{L}\\ \label{eq:1.12}
\ee
\be
\bigtriangleup \hat{R}'  =  -2G \hat{E} \label{eq:1.13}
\ee
where $\bigtriangleup \hat{\Pi}_L =\lim_{\epsilon\to 0^+}
\Pi_L (\hat{r} + \epsilon) - \Pi_L
(\hat{r} - \epsilon)$, etc.\\
The quantities $L(r)$ and $R(r)$ represent the gauge degrees of freedom whereas
$\hat{r}$ represents the dynamical degree of freedom.  The
arbitrariness of $L(r)$ and $R(r)$ represents the reparameterization
invariance of classical gravity.  The gauge fixing to determine the
radial parameterization will be considered in section 4.

\section{QUANTIZATION}

Having solved the constraints, we can proceed with Dirac quantization.
However, we will ignore the quantum ordering problem \cite{henneaux}.
The Poisson
brackets are now replaced with commutators for quantum operators;
e.g., $[L^{op}(r'), \Pi_L^{op}(r)] = i\hbar\delta(r'-r)$.  In the
coordinate representation, $\Pi_L^{op} \rightarrow (\hbar/i) \delta/\delta
L$, where $\delta f/ \delta L$ represents the functional derivative of
functional, $f$.  Therefore, the wave function, $\psi (L,R, \hat{r})$
is the solution of the following equations
\be
(\hbar/i)\delta \psi/ \delta L = \Pi_l \psi;\;\;\;\;(\hbar/i) \delta \psi/
\delta R = \Pi_R \psi \label{eq:3.1}
\ee
where $\Pi_L$ and $\Pi_R$ are given by \req{eq:1.10} and \req{eq:1.11}.  Let
\be
\psi = \exp(iS/t).  \label{eq:3.2}
\ee
Then \req{eq:3.1} reduce to
\be
\delta S/ \delta L = \Pi_L;\;\;\;\;\delta S/ \delta R = \Pi_R. \label{eq:3.3}
\ee
The following solution of the above simultaneous equation was found,
which can be verified by a direct substitution:
\be
S  =  \int_0^{\hat{r}} drS_{-}  + \int_{\hat{r}}^\infty drS_+;
 \label{eq:3.4}
\ee
where
\be
S_\pm
:= \frac{1}{4G}\left[ Q_\pm + \frac{R'}{2}\ln{\frac{Q_\pm-R'}{Q_\pm+R'}}
\right]_{\pm},
\ee
with
\be
Q_+ :=  [(R'/L)^2 + 8GM]^{1/2};\;\;\;\;Q_{-} := [(R'/L)^2 + \lambda
R^2]^{1/2}. \label{eq:3.5}
\ee
Here $+[-]$ refer respectively to the regions $r > \hat{r} [r < \hat{r}]$.

It follows from \req{eq:3.5} that $S_+$ is always real and $S_{-}$ is real if
$\lambda > 0$ (de Sitter space-time).  In the latter case, the probability
density
\be
P = \psi^* \psi, \label{eq:3.6}
\ee
is a constant ($=$ 1, ) independent of
$L,\;\;R\;\;\mbox{or}\;\;\hat{r}$.  This represents an unbounded motion
where the string has the same probability for any radius,
$\hat{r}$.  However bound states can exist if $\lambda < 0$ (anti-de
Sitter space time).  In such a case one or more classical
turning points may exist, whenever
\be
R'(r)^2 = - \lambda [R(r) L(r)]^2. \label{eq:3.7}
\ee
The bound states will be analysed in the next section.

The wave functional $\psi$ in \req{eq:3.2} is in WKB form.  This is due to
the operator ordering implied by \req{eq:3.1}, which is essentially
equivalent to a WKB approximation with arbitrary $L(r)$ and $R(r)$.
\req{eq:3.2} represents an infinite set of wave functionals for pure gravity.
However, if one fixes the gauge to satisfy the boundary condition,
which represent the coupling to the string, a particular wave {\it function}
can be found for the whole system.  This will be done in the next
section.

\section{GAUGE FIXING AND COUPLING TO THE STRING}
Here we will specialize to the case of an anti-de Sitter space-time in
the inner region $(r < \hat{r})$.  In the absence of the string, the
appropriate choice of gauge would be
\bea
R(r) & = & r \nonumber \\
L(r) & = & \left\{ \begin{array}{ll}
        1, & r > \hat{r}\\
        (1-\lambda r^2)^{-1/2}, & r < \hat{r}
\end{array} \right. \label{eq:4.1}
\eea
We will consider a two parameter gauge which modifies the above
space-time such that the parameters are completely determined by the
boundary conditions.  An appropriate choice is
\bea
R(r) & = & r[1 - \beta \exp(-\alpha|r-\hat{r}|)] \nonumber \\
L(r) & = & \left\{ \begin{array}{ll}
   1 - \beta \exp[-\alpha (r - \hat{r})], & r > \hat{r}\\
  (1 - \beta)(1 - \lambda \gamma^2)^{-1/2} + \beta \exp[\alpha(r -
\hat{r})], & r < \hat{r}
\end{array} \right. \label{eq:4.2}
\eea
where $\alpha$ is positive and it represents the range of the coupling
to the string, and $\beta$ represents the strength of that coupling.
Note that
\req{eq:4.2} satisfies the limiting behaviour, \req{eq:4.1}, as $r \rightarrow
\infty$ or as $\beta \rightarrow 0$.

First we determine $\beta$ using the continuity of the metric at $r =
\hat{r}$.  $R(r)$ is already continuous, and the continuity of $L(r)$
gives
\be
\beta = [1-(1-\lambda \hat{r}^2)^{1/2}][1-2(1-\lambda
\hat{r}^2)^{1/2}]^{-1}. \label{eq:4.3}
\ee
Note that $\beta \rightarrow 0$ if $-\lambda \hat{r}^2 \rightarrow 0$
and $\beta \rightarrow 1/2$ if $-\lambda \hat{r}^2 \rightarrow
\infty$.  Next, we combine the boundary conditions \req{eq:1.12} and
\req{eq:1.13}
by eliminating $\hat{p}$.  The result is
\be
(\bigtriangleup \hat{R}')^2 = 4G^2 \hat{L}^2[4(\bigtriangleup
\hat{\Pi}_L)^2 + \mu^2 \hat{R}^2]. \label{eq:4.4}
\ee
Substitution of \req{eq:4.2} into \req{eq:4.4} gives an algebraic equation for
$\alpha$ in terms of $\hat{r}$ and $\beta(\hat{r})$.
Note that $\alpha$ and $\beta$ are completely determined by the string
parameters $\hat r$ and $\mu$.
An examination
of limiting cases can reveal the main qualitative features of the
classical motion implied by \req{eq:4.4}.

First consider the case with a small mass and cosmological constant, so that
$4MG/(R'/L)^2 << 1$ and $\lambda R^2/(R'/L)^2 << 1$.  Note that $R'/L =
1$ for a flat space.)  In such a case \req{eq:4.4} reduces to a quadratic
equation.  We are looking for a real positive root which is given by
\be
\alpha = [-b - (b^2 - 4ac)^{1/2}](2a)^{-1}, \label{eq:4.5}
\ee
when $a = -4[4MG+2 \lambda \hat{r}^2 (1-\beta)^2 + G^2 \mu^2
\hat{r}^2]\beta^2 \hat{r},\;\;b = 4[4MG - 2 \lambda \hat{r}^2
(1-\beta)^2](1- \beta) \beta,\;\;\mbox{and}\;\;c = 4G^2b^2(1-\beta)^2
\hat{r}$.  If the $\lambda \hat{r}^2$ term is small, $a < 0,\;\;b >
0,\;\;c > 0$.  Therefore $\alpha \geq 0$ in such a case.  To take a
closer look at the situation, consider the case where the string
tension is dominant.  Then \req{eq:4.5} reduces to
\be
\alpha \hat{r} = \beta(1-\beta) \approx -2/ \lambda \hat{r}^2 >>
1. \label{eq:4.6}
\ee
That is, the distortion of the space-time produced by the string has a
very short range.

According to \req{eq:3.7}, there is no classical turning point in the limit
considered above.  The more interesting case of a bound state occurs
if there is a classical turning point at $r = r_0$; this is possible
if $r_0 < \hat{r}$.  If we further assume that $r_0$ is close to
$\hat{r}$, then \req{eq:4.4} has a single root given by
\bea
\alpha & = & \frac{1 - \beta}{\beta} [(G^2 \mu^2 -
\lambda/4)(1-\beta)^2 \hat{r}^2 + 2GM]^{1/2}\\ \label{eq:4.7}
r_0 & = & ( \sqrt{- \lambda} + G \mu)^{-1}. \label{eq:4.8}
\eea
Again one obtains \req{eq:4.6} if the tension term is dominant in \req{eq:4.7}.
Therefore, \req{eq:4.6} gives a reasonable estimate of the range of
interaction for a small cosmological constant.  (There is also the
possibility of multiple classical turning points, with a  probability
of quantum tunneling between different classical regions.  We will not
consider such cases here.)

Returning to the quantum behaviour of the system, it follows from
\req{eq:3.6} that
\be
P = \exp[-2 Im\left( S/\hbar\right)], \label{eq:4.9}
\ee
gives the quantum probability of classical barrier penetration.  As
noted earlier, $Im S$ can only result from the term in $S_-$.  Separating
the real and imaginary parts in \req{eq:3.4}, we obtain for the case of one
classical turning point, $r_0$.
\be
S_{-} = \frac{1}{4G} \int_0^{r_0} dr \left[ Q_{-} -R' \ln \frac{Q_{-} +
R'}{\sqrt{- \lambda} R L} \right] + \frac{i}{4G} \int_{r_0}^{\hat{r}}
dr \left[ \bar{Q} - R' \cos^{-1} \left( \frac{R'}{\sqrt{- \lambda} R
L} \right) \right], \label{eq:4.10}
\ee
where $\bar{Q} := (- \lambda R^2 L^2 - R'^2)^{1/2}$.  For small $-
\lambda$, \req{eq:4.9} and \req{eq:4.10} give
\be
P(\hat{r}) = \exp[-(1 - \beta) \sqrt{\lambda} \hat{r} (\hat{r}^2 -
r_0^2)/4 \ell_0^2], \label{eq:4.11}
\ee
where $\hat{r} \ell_0^2 = \hbar G$ with $\ell_0 =$ Planck length \cite{foot}.
Therefore,
the barrier is determined by Planck length, which is
characteristic of quantum gravity.  The resulting uncertainty in the
radius of the string, $\bigtriangleup \hat{r}$, is easy to obtain from
\req{eq:4.1}.
\be
\bigtriangleup r \approx (2/\sqrt{- \lambda}) (\ell_0/\hat{r})^2.
\label{eq:4.12}
\ee
It follows that $\bigtriangleup r << (- \lambda)^{-1/2}$ if $\hat{r}
>> \ell_0$, but $\bigtriangleup \hat{r} \sim (- \lambda)^{-1/2} >>
\hat{r}$ if $\hat{r} \sim \ell_0$.  That is, the radius of a string is
completely uncertain in this limit.  This is to be expected, since the
quantum fluctuation of space-time are expected to be large on a Planck
scale.

The form chosen for the radial parameterization, \req{eq:4.2}, is quite
appropriate for the simple distortion of space-time considered here.
The general solution is valid for other choices of gauge as long as
the boundary conditions are satisfied.  Of course, the complete gauge
fixing also requires the choice of ``time slice'' to define the
dynamics.  In the spirit of this work, this can be done at the
classical level by solving the classical equation of motion that
follow from \req{eq:1.4}.  However, this is a separate problem which is
outside the scope of this work.

\section{CONCLUSIONS}

We have considered some new features of the problem of a circular
string, which form the boundary of a two component space-time.  Unlike
the previous works \cite{soda}, we assume an active role for the string in the
distortion of background space-time.  The model is simple enough so that the
constraints can be explicitly solved, and one can proceed with Dirac
quantization.  The general wave functional, $\psi(L(r), R(r),
\hat{r})$, is valid for arbitrary gauge choice as long as the boundary
conditions that represent the coupling to the string are satisfied.
This is illustrated with a two-parameter gauge fixing; it leads to a
$\psi$ which depends only on the true dynamical degree of freedom,
$\hat{r}$.  The quantum behaviour of the system is examined in some
limiting cases.  In particular, the quantum uncertainty in the radius
of the ring, $\bigtriangleup \hat{r}$, is calculated in the limit of
large tension in the string.  If $\hat{r} \sim \ell_0$, it is found
that $\bigtriangleup \hat{r} >> \hat{r}$, i.e. a very ``fuzzy''
string.  As expected, the quantum fluctuations of space-time are large
in this limit, and the notion of a sharp circular string breaks down.

There are several interesting aspects of the problem which require
further work.  For example, it is possible that more than one
classical turning points exist for our choice of gauge, or for other
choices.  In that case quantum tunneling could occur between different
classical regions.  In addition, the choice of ``time slice'' to study
the time evolution of the system should be interesting.  As noted
earlier, this will require a solution of classical equations of
motion.  In such a case, it should be possible to obtain a canonical
transformation \cite{carlip2} relating the time-dependent results to our time
independent results.  However, our immediate interest lies in the
first order approach \cite{review,witten1,ash1},
where the emphasis is on the topological
aspects.  This is the subject of the following paper.

\bigskip\noindent
{\bf Acknowledgments}

\noindent
The authors wish to thank D.E. Vincent for his participation in the early
stages of this work.  We acknowledge the partial support of the Natural
Sciences and Engineering Research Council of Canada.
\end{document}